# Deformation of an inflated bicycle tire when loaded

Jordi Renart[1] and Pere Roura-Grabulosa[2*]

[1]AMADE, Polytechnic School, University of Girona, Campus Montilivi s/n, E-17003 Girona, Spain.

[2] GRMT, Department of Physics, University of Girona, Campus Montilivi, Edif.PII, 17003-Girona, Catalonia, Spain.

*Corresponding author: pere.roura@udg.cat

**Abstract**

The deformation of a loaded bike tire has been analyzed with a model consisting of a toroid of thin inextensible walls mounted on a central rim. If the tire radius is much shorter than the rim radius, the deflection, d, of the tire can be calculated as a function of the applied load, F. The solution can be approximated to a power law dependence $F \propto d^{3/2}$ for small loads. The theoretical predictions compare well with the experiments carried out on two bicycle tires.

## 1.Introduction

Contact Mechanics deals with the deformation in the contact region of two rigid bodies that are pushed together. H.Hertz was the first to solve this kind of problems and the so-called Hertz theory is restricted to solid bodies in the linear-elastic regime. Exact solutions for many specific cases such as a sphere or a conical indenter on the flat surface of a semiinfinite body, or two parallel or perpendicular cylinders pressed together can be found in the classical literature on elasticity.[1,2] Since, due to local deformation, the contact area increases with the applied force, F, a supralinear dependence like

$$F = k \cdot d^n \quad \text{with } n > 1 \tag{1}$$

is always found where d is the relative displacement of the two bodies and k collects all the geometrical factors and elastic constants.

The exact analysis of these kinds of problems requires an in-depth knowledge of elasticity and mathematical methods that is beyond the abilities of undergraduate students.

However, sound approximations that retain the essential physics can be applied to specific problems. For instance, that of two identical spheres which was solved by B. Leroy with remarkable accuracy.[3]

The aim of this paper is to analyze the static deformation of a pneumatic tire when loaded on a flat surface. Our interest was triggered by a simple paradox. The tire holds the rim (and, consequently, the load applied to the wheel axis) thanks to pressure from the inner gas. However, since the pressure is the same at any point, its net vertical force on the rim will be null. Of course, the reader will claim that we have missed the action resulting from tire deformation, and this will be the contents of our paper.

The geometry of a real tire and its construction is very complicated.[4,5] The casing resembles a toroid but with varying wall thickness and it has reinforcing cords embedded in the rubber at particular directions. Consequently, analyzing its mechanical behavior (static and dynamic) mostly relies on finite element calculations and parametric approaches to organize the experimental results observed. When searching through the specialized literature on this subject,[4-7] the authors were surprised by the lack of simple models to help attain an initial elementary quantitative comprehension of the statics of a pneumatic tire under load.

We have, thus, chosen the simplest model consisting of a toroidal inextensible membrane "truncated" by a central rim. In addition, we consider that the toroid is "thin" compared to the wheel radius. Under these assumptions, the geometry of the deformed tire can be obtained by numerically solving a system of algebraic equations and the displacement-load curve can thus be predicted. An interesting point is that, under a reasonable assumption about the geometry of the contact patch with the ground, we obtain a power dependency like that of Eq.(1) valid for small deformations. The theoretical predictions are then tested satisfactorily against experiments carried out on two bicycle wheels. The paper finishes with a concluding section which summarises the results.

## 2.Theoretical section

The paradox raised in the Introduction has a simple qualitative solution.[8,9] The fact is that, in addition to the air pressure and applied load, the rim will experience an upward force from the tire. Let us have a look at this force. When unloaded, the tire is symmetric around the wheel axis. Due to the air pressure it is deformed and a biaxial tension, $\sigma$, arises on its surface. The force per unit length at the contact line with the rim will be the same at any point, and no net force will result (Fig.1a). However, when loaded, the tire will become deformed near the ground (Fig.1b). Its cross section will be flattened and the vertical component of the force exerted on the bottom of the rim will be lower than on the upper part (because of the new angle φ – see Fig.1b – and because of the tension reduction $\sigma_L < \sigma$ – see below). Consequently, when integrated all around the rim, the tire will exert a net upward force that will equilibrate the load, F. We can thus say that tire sustains the load through the biaxial tension created by the air pressure, notably, at the upper half of the tire.[8] When poor inflated, the lowest part of the tire can also contribute to the upward force (Fig.1c).

This section will be devoted to develop quantitatively this analysis to predict the relationship between the applied load and the downward displacement of the wheel.

### *2.a Biaxial stress on a toroidal tire*

Consider an inflated toroidal tire whose dimensions are defined by radii $R_W$ and $R_{T0}$ (Fig.2), with the condition $R_W >> R_{T0}$. If its thickness, h, is small enough, it can be considered a membrane, which can only hold tensile stresses. In this case, the internal air pressure, P, will be sustained exclusively by the two principal stress components $\sigma_1$ and $\sigma_2$ acting on the tire surface (Fig.2). At any point, Young-Laplace's equation:[10]

$$P = h\left(\frac{\sigma_1}{R_1} + \frac{\sigma_2}{R_2}\right), \tag{2}$$

where $R_i$ are the principal radii of curvature, must be obeyed. For a "thin" toroid ($R_W >> R_{T0}$), at any point $R_1 << R_2$. Along the particular line defined by $R_W$ (Fig.2), $R_2$ is infinite and, application of Eq.(2) leads to:

$$\sigma_1 = \frac{R_{T0}P}{h} \tag{3}$$

It can be shown (Appendix I) that the value along this line is the same than the average value over all the tire's surface, $\bar{\sigma}_1$. In fact, since $\sigma_2$ is always positive, Young-Laplace's equation tells us that $\sigma_1 < \bar{\sigma}_1$ beyond $R_W$ ($R_2$ is positive there) whereas the contrary holds for points inside $R_W$ (where $R_2$ is negative). We will consider that $\sigma_1$ can be approximated by $\bar{\sigma}_1$, this assumption being accurate since $R_W >> R_{T0}$.

In Appendix I, more accurate values of $\sigma_1$ and $\sigma_2$ are derived.

### *2.b Wheel geometry before loading*

A wheel consists of an inflated tire mounted around a rigid rim of radius $R_L$ and width $W_L$ (Fig.1a). We will consider that the tire acquires the shape of a toroid "truncated" by the rim.

Before loading, the tire cross section can be described by several sets of two independent parameters, the most natural being $W_L$ and $R_{T0}$. It is easy to realize that, since $W_L$ and $R_{T0}$ determine the tire contour (i.e. the path length of the cross section), C, the tire "height", $H_0$, and the angle, $\varphi_0$ at the point of contact with the rim (Fig.3b), according to equations:

$$cos\varphi_0 = \frac{W_L}{2R_{T0}}$$

$$H_0 = R_W + R_{T0} - R_L$$

$$C = W_L + R_{T0}(2\varphi_0 + \pi), \qquad (4)$$

any pair of parameters chosen among $W_L$, $R_{T0}$, C, $H_0$ and $\varphi_0$ is enough to describe the tire cross section.

### *2.c Geometry of a loaded wheel*

When a load F is applied to the wheel hub, the tire cross section will be "compressed", i.e., its "height" measured along a radial direction defined by angle θ will diminish and the tire will become flat in contact with the ground along a length $W_C(\theta)$ (Fig.3c). Since we model the tire as an inextensible membrane, the only parameters that will remain unaffected by the deformation will be the contour C and $W_L$. If we assume that the curved part of the cross section has constant curvature (see below), then the deformed cross section can be described with $W_L$, C and one parameter among $R_T(\theta)$, $\varphi(\theta)$ and $H(\theta)$. The aim of this subsection is to describe the tire cross section at any radial direction when, due to the load, the wheel hub is displaced downward by d(0). From now on, and according to the literature, d(0) will be called the tire "deflection".

To predict the geometry of the loaded tire and its load-deflection curve we assume two approximations. The first one is that the curved part of the cross-section is circular with constant radius of curvature, $R_T(\theta)$, along any radial cross section (Fig.3c). This is equivalent to assume that the tire pressure is held by the tension along this direction, the contribution of the other principal direction being negligible. Similarly to the case of the toroid, for $R_{T0} << R_W$ this is reasonable. The second approximation concerns the angle $\theta_C$ where the tire loses its contact with the ground. As illustrated in Fig.3a, we assume that its value is:

$$cos\theta_C = 1 - \frac{d(0)}{R_W + R_{T0}}. \qquad (5)$$

For $\theta > \theta_C$, the tire will remain undeformed; i.e. its cross section will be the same irrespectively of the applied load. In this region, $\varphi_0$, $H_0$ and C are given by the set of Eqs.(4).

For $\theta < \theta_C$, the tire cross section will become flat in contact with the ground and, consequently, its curved part will have a radius of curvature $R_T(\theta)$ smaller than $R_{T0}$ and a higher contact angle with the rim, $\varphi(\theta)$ (Fig.3c). Now the contour will be:

$$C = W_L + R_{T0}[2\varphi(\theta) + \pi] + W_C(\theta), \qquad (6)$$

where $W_C(\theta)$ is the contact length of the tire with the ground (Fig.3c):

$$W_C(\theta) = W_L - 2R_T(\theta)cos\varphi(\theta). \qquad (7)$$

Introduction of Eq.(7) in Eq.(6) leads to:

$$C = 2W_L + R_T(\theta)[2\varphi(\theta) + \pi - 2cos\varphi(\theta)]. \qquad (8)$$

The wheel geometry will be perfectly defined if $\varphi$ and $R_T$ are known for any value of $\theta$ as a function of the deflection, d(0), and the unloaded wheel geometry. Eq.(8) can be reduced into one equation on only $\varphi(\theta)$ if we introduce the tire "height" $H(\theta)$ defined in Fig.3c:

$$R_T(\theta) = \frac{H(\theta)}{1+sin\varphi(\theta)}, \qquad (9)$$

Substitution of Eq.(9) in Eq.(8), leads to:

$$\frac{\pi+2\varphi(\theta)-2cos\varphi(\theta)}{1+sin\varphi(\theta)} = \frac{C-2W_L}{H(\theta)}. \qquad (10)$$

In summary, numerical solution of Eq.(10) delivers $\varphi(\theta)$ and Eq.(9), $R_T(\theta)$. The geometry of the loaded wheel is thus determined once $H(\theta)$ is known. Due to the load, the distance from the center of the wheel to the ground along the $\theta$ direction is reduced by the amount (Fig.3a):

$$d(\theta) = -\frac{R_W+R_{T0}-d(0)}{cos\theta} + R_W + R_{T0}\ . \qquad (11)$$

and $H(\theta)$ is simply:

$$H(\theta) = R_W + R_{T0} - R_L - d(\theta) \equiv H_0 - d(\theta). \qquad (12)$$

### *2.d Deflection-load dependence*

Once the tire geometry of the loaded wheel has been established, the dependence of the load on deflection d(0) can be derived. As already commented on above, the load is equilibrated thanks to the tire deformation. This condition can be written as:

$$F = -2R_L h \int_0^{2\pi} \sigma_1(\theta)cos\varphi(\theta)cos\theta d\theta. \qquad (13)$$

where the variation of h with $\theta$ has been neglected, in agreement with the assumed constancy of the contour C. Since the resultant of the tire tension is null for the unloaded wheel we can subtract it from Eq.(13):

$$F = -2R_L h \int_0^{2\pi}[\sigma_1(\theta)cos\varphi(\theta) - \overline{\sigma}_1 cos\varphi_0]cos\theta d\theta, \qquad (14)$$

that, finally can be simplified as:

$$F = -4R_L h \int_0^{\theta_c} [\sigma_1(\theta) cos\varphi(\theta) - \overline{\sigma}_1 cos\varphi_0] cos\theta d\theta, \qquad (15)$$

because, beyond $\pm\theta_C$, $\sigma_1(\theta) = \overline{\sigma_1}$, and $\varphi(\theta) = \varphi_0$. Finally, application of Young-Laplace's Eq.(3), leads to the desired result:

$$F = 4R_L P \int_0^{\theta_c} [R_{T0} cos\varphi_0 - R_T(\theta) cos\varphi(\theta)] cos\theta d\theta, \qquad (16)$$

For a particular deflection, d(0), $R_T(\theta)$ and $\varphi(\theta)$ can be obtained through the solution of Eqs.(9) and (10) and the load F can be calculated after numerical integration of Eq.(16). These equations have been solved by MATLAB software package. A typical F *vs* d(0) curve obtained from the particular values of a road-bike wheel has been plotted in Fig.4a.

***2.e Small-deflection limit***

The visual inspection of Fig.4a shows that the relationship is non-linear and that it begins with zero slope. In the inset of Fig.4a, the log-log plot reveals that, for small deflection:

$$F(N) = 44.16\, d(0)^{\frac{3}{2}} \qquad \text{[d(0) in mm]} \qquad (17)$$

The simple value of the power poses the challenge of deriving it from the deflection-load dependence of Eq.(16), that was obtained after quantification of the tire stress action along the rim border. Since the functional dependence on d(0) cannot be made explicit from Eq.(16), we decided to look for an alternative way to calculate F.

Since a membrane cannot sustain bending forces, the membrane contour is tangent to the ground and so cannot exert any upward force. The force per unit area exerted by the tire on the ground will be constant and equal to the gas pressure, P. Consequently, the force sustaining the wheel will be equal to the footprint area, A, times P:

$$F = P \cdot A\ , \qquad (18)$$

where A can be calculated because the footprint width is known as a function of θ [$W_C(\theta)$ in Eq.(7)]. With the approximation $R_W + R_T(\theta) \approx R_W + R_{T0}$, it is easy to deduce that

$$A = -(R_W + R_{T0}) \int_{-\theta c}^{\theta c} R_T(\theta) cos\varphi(\theta) cos\theta d\theta, \qquad (19)$$

and, following steps similar to those of the former section, we arrive at

$$F = 4(R_W + R_{T0}) P \int_0^{\theta_c} [R_{T0} cos\varphi_0 - R_T(\theta) cos\varphi(\theta)] cos\theta d\theta, \qquad (20)$$

that coincides with Eq.(16) except for the substitution of $R_L$ by $R_W + R_{T0}$. In the limit of $R_{T0} \ll R_L, R_W$, the relative discrepancy is $2\, R_{T0}/ R_W$. Its origin should be found in the assumption that $\sigma_2$ does not contribute appreciably to the value of P and, consequently, that $\sigma_2$ is constant leading to sidewall arcs of the tire cross section of constant curvature.

For our purpose of deducing the power 3/2, the procedure involving the footprint area is very suitable. We can reasonably assume that the footprint is bound between that of a rhomb whose diagonal lengths are $W_C(0)$ and $2L_C$:

$$2L_C = 2(R_{T0} + R_W)sin\theta_C \qquad (21)$$

and that of an ellipse of semi axes $W_C(0)/2$ and $L_C$. Consequently,

$$F = \alpha P W_C(0) L_C \qquad \text{where} \quad 1 \text{ (rhomb)} < \alpha < \pi/2 \text{ (ellipse)}. \qquad (22)$$

In Appendix II we show that:

$$F = \alpha\beta P d(0)^{3/2}, \qquad (23)$$

where $1 < \alpha < \pi/2$ between the limits of rhomboidal and elliptical footprints, respectively, and $\beta$ is a constant that depends on the geometry of the unloaded wheel. The footprints of the loading curve of Fig.4a at several loads are shown in Fig.4b. In any condition, the footprint area is larger than that of a rhomb. Consequently we expect $\alpha$ to be larger than 1.

As shown in Appendix II, $\beta$ is the product of two factors:

$$\beta = \sqrt{2(R_W + R_{T0})} \cdot \left(\frac{b}{a}\right) \quad . \qquad (24)$$

The first factor depends on the wheel radius, i.e., its size along the plane of the wheel. The second factor depends on the cross section of the undeformed tire. A priori, one would expect a dependence on the shape and size of this cross section. The surprising result is that it depends exclusively on the angle of contact $\varphi_0$ (Fig.5):

$$\left(\frac{b}{a}\right) = \frac{2cos\varphi_0}{1+sin\varphi_0} - 2\frac{\pi+2\varphi_0-2cos\varphi_0}{cos\varphi_0(\pi+2\varphi_0)-4(1+sin\varphi_0)} \qquad (25)$$

that, according to Eq.(4), is a function of the ratio $W_L/R_{T0}$. In other words, the load-deflection curve depends on the shape but not on the size of the tire cross section.

Application of Eqs.(23), (24) and (25) to the curve of Fig.4a requires introduction of an additional factor in Eq.(23) equal to $R_L/(R_W+R_{T0}) = 0.904$ to take into account the discrepancy between the value of F deduced from the stress around the rim [Eq.(16)] and from the footprint area Eq.(20). With this correction, the value of $\alpha$ is 1.259.

We are now ready to analyze the goodness of the power law dependency to describe analytical curve of Fig.4a. For a load of 659 N, which would correspond to the load on the rear wheel when a total weight of 104 kg is supported by a bicycle with a front/rear weight distribution of 35/65, the deflection is 6 mm. If the power law curve is used, instead, for the same deflection the load is only 10% higher, meaning that this approximation is valid for realistic loading conditions (the prediction imposing $\alpha = 1$ would lead to a value 5% lower). Finally, the cross section at the bottom of the tire has been calculated and drawn in Fig.4c.

Before leaving this section, we should insist on the fact that, with identical approximations, the two approaches to calculate F (from stress and from pressure) should lead to identical results if the tire behaves like a membrane. Real tires are far from a membrane. As a result, the force per unit area is not constant at the footprint[4] and application of Eq.(18), with P being the gas pressure, always overestimates the load.[11]

## 3. Experimental section

### *3.a The wheels*

Since the model consists of a truncated toroidal membrane, we first tried to compare the prediction with the behavior of the inner tube of a wheel mounted on its rim. Preliminary tests showed us that it was unsuitable because the tube rubber was far from inextensible; when loaded its contour at the point of contact with the ground diminished drastically (incidentally, we can say that this is in agreement with Young-Laplace's Eq.(3)). We thus decided to do the experiments without taking out the casing. In this case, the most apparent departure from the model was the thread pattern. However, another important feature was the variation of the wall thickness below the pattern. To avoid these deviations from the model the tire was ground until its thickness was constant. This grinding operation was done at the two opposite points of the wheel that took contact with the compression plates used for the tests.

We did the experiments with a mountain-bike and a road-bike wheel. Their dimensions ($R_L$, $R_W$, $W_L$ and $R_{T0}$) are detailed in Figs.6 and 7 (both tires are nearly inextensible: when the gas pressure was increased by 2.5 bar, $R_{T0}$ increased by only 3%). After grinding, the tire thickness was 2.1±0.1 and 2.2±0.1 mm for the mountain-bike and the road-bike tires, respectively.

### *3.b The tests*

The deformation experiments were done with a universal machine (MTS Insinght 100 kN). A displacement was applied to the wheel by means of two compression plates (Fig.8) and the load was recorded with a 1 kN load cell.

Before the test, the tire was inflated to the desired pressure. Then the wheel was located between the compression plates and an initial 5 N load was applied to hold it. The verticality was ensured with a bubble level.

The tire was compressed at a constant rate of 5 mm/min until the load reached 500 N. In order to test the repeatability of the measurements, the test was run three times for each pressure level. Between the tests, the tire was not removed from the machine.

The tests have been done with a commercial apparatus because it is available in our school for mechanical testing. However, accurate measurements can also be done with a very cheap home-made set up consisting of a rigid structure with a screw on its top to compress the tire and a load cell in between.

### *3.c Comparison between experimental and predicted load curves*

The measured curves are plotted in Figs.6 and 7. As expected from general experience, the displacement at any particular load is higher when de tire is inflated at a lower pressure. The non-linear dependence is also significant and qualitatively agrees with our theoretical analysis. However, the crucial point is to assess to what extend our simple model of tire deformation can quantitatively predict the tire behavior. To achieve this aim, several points have to be considered. First, since the load was not applied to the wheel hub but between the top and the bottom of the wheel (Fig. 8), the measured displacement, D, is the result of the deflection at these two points. Second, the wheel weight, mg, produces a larger deflection at the bottom. And finally, at the preload $F_0$ = 5 N, the displacement recorded by the machine was zero.

For a given pressure, the theoretical load vs deflection curve was calculated for the particular wheel. In fact, we are interested in the inverse function, d(F). Once this curve is known, the experimental displacement, D(F) can be calculated according to:

$$D(F) = d_b(F + mg) + d_t(F) - d_b(F_0 + mg) - d_t(F_0) \quad , \qquad (26)$$

where $d_b$ is the deflection at the bottom that takes into account the wheel weight, $d_t$ is the deflection at the top and $d_b(F_0+mg)$ and $d_t(F_0)$ are the deflections when only the preload is applied.

The predictions are plotted as solid lines in Figs.6 and 7. Both the shape of the curves and the general evolution when pressure is increased are fairly well predicted. Deviations between the model and the experiment are systematic but have opposite signs in both experiments. This fact makes it very difficult to elucidate their origin. On the other hand, with the exception of the curve measured at 6 bar, the predicted displacement at 500 N matches the experimental value with an error below 5 %. This is a value similar to the accuracy of the analytical prediction, which is estimated to be around $R_{T0}/R_W$ (6-7%).

## 4. Summary

The mechanical deformation of a tire has been analyzed. The gas pressure induces a biaxial tension that, when the tire is loaded, becomes lower at the bottom of the tire resulting in net upward force acting on the rim.

For a simple but realistic model of a bicycle tire, the load-deflection curve has been predicted. The model consists of an inextensible thin membrane of toroidal shape that is truncated by a cylindrical rim. Since we have assumed that the tire radius is much shorter than the wheel radius, the tension along the direction of the wheel contour can be neglected and the geometry of the loaded tire can be described by a system of algebraic equations. For any radial direction, the tire cross section has a constant radius of curvature. This curvature radius reaches a minimum value at the tire bottom and increases steadily until the tire loses its contact with the ground. Once the geometry is known, the upward force that the tire exerts on the rim can be calculated by integration. The result is a non-linear dependence that for small deflection can be described by a power law with exponent 3/2. This exponent has been successfully predicted following an approximate alternative method. The upward force has been calculated as pressure times the area of contact with the ground. For a given (small) deflection, it has been shown that F depends on the wheel radius but not on the tire radius.

The predicted load-deflection curves have been compared with experimental curves measured on two commercial bicycle wheels. From the fair agreement achieved we conclude that the proposed model has been satisfactorily validated.

**Acknowledgments**

This work has been partially funded by the MPCUdG2016 program of the University of Girona. The authors are indebted to Dr.Lluís Ripoll for exposing to them the paradox that is at the origin of this study.

## Appendix I. The value of $\sigma_1$ and $\sigma_2$

Consider half of the toroidal tire once “cut” through the torus plane (Fig.9). Let us analyze the mechanical equilibrium of forces acting on the portion of the tire that is farther than $R_W$ from its axis of symmetry ($r > R_W$). This portion remains in equilibrium because the force due to pressure is equal to the net force arising from the stress at the border of this tire portion. The force due to pressure is horizontal and of magnitude:

$$F_{Pe} = \pi(R_e^2 - R_W^2)P \ , \tag{I.1}$$

where $R_e \equiv R_W + R_{T0}$. Concerning the stress, since we consider that the tire is a perfect membrane, it can be only under tension. At the border placed at r = $R_W$, the force due to stress is radial at any point and, consequently, after integration cancels out. At the border at r = $R_e$, the tension along the principal axis 1 (Fig.2), $\sigma_{1e}$, leads to a horizontal net force:

$$F_{\sigma 1e} = 2\pi h R_e \sigma_{1e}. \quad \text{(I.2)}$$

The equality $F_{Pe} = F_{\sigma 1e}$ leads to the exact value of $\sigma_{1e}$:

$$\sigma_{1e} = \frac{P}{2h}\frac{2R_W R_{T0} + R_{T0}^2}{R_W + R_{T0}}. \quad \text{(I.3)}$$

The same analysis applied to the torus cross section "inside" $R_W$ delivers the exact value of $\sigma_1$ at r = $R_i \equiv R_W - R_{T0}$:

$$\sigma_{1i} = \frac{P}{2h}\frac{2R_W R_{T0} - R_{T0}^2}{R_W - R_{T0}} \quad . \quad \text{(I.4)}$$

If we expand these formulae up to the first order in $R_{T0}/R_W$, we obtain:

$$\sigma_{1e} \approx \overline{\sigma}_1 \left(1 - \frac{1}{2}\frac{R_{T0}}{R_W}\right) \quad , \quad \text{(I.5)}$$

and

$$\sigma_{1i} \approx \overline{\sigma}_1 \left(1 + \frac{1}{2}\frac{R_{T0}}{R_W}\right) \quad , \quad \text{(I.6)}$$

where $\overline{\sigma}_1 \equiv \frac{R_{T0}P}{h}$ can be considered as the average value of $\sigma_1$ that coincides with the exact value at r = $R_W$ derived in Section 2a. [Eq.(3)]. The values of $\sigma_1$ derived here at $R_i$ [Eq.(I.3)], $R_W$ [Eq.(3)] and $R_e$ [Eq.(I.4)] coincide with those calculated by other authors[12] for torus of $R_{T0}/R_W$ =1/1.5.

The average value of the other principal stress, $\overline{\sigma}_2$, can be obtained if a radial cross section of the tire is considered instead (Fig.9b). Mechanical equilibrium leads to

$$\overline{\sigma}_2 = \frac{R_{T0}P}{2h} = \frac{\overline{\sigma}_1}{2}. \quad \text{(I.7)}$$

We can go beyond this average value and calculate the exact value of r = $R_e$, $R_w$ and $R_i$. Substitution the corresponding values of $\sigma_1$ into Young-Laplace's Eq.(2) gives a surprising constant value of $\sigma_2$ equal to $\overline{\sigma}_2$ that agrees with the result given by the linear theory.[12]

**Appendix II. The power law for a small applied load**

If the load, F, has a power dependence on deflection, d(0), (Fig.4) it is natural to expect that it comes from similar dependencies of $W_C(0)$ and $L_C$ [see Eq.(II.4)], i.e. we assume that:

$$W_C(0) \propto d(0)^n \qquad \text{and} \qquad L_C \propto d(0)^m, \quad \text{(II.1)}$$

for d(0) small (small load).

Introduction of Eq.(5) into Eq.(21) leads to:

$$L_C = \sqrt{2(R_{T0} + R_W)}d(0)^{1/2} \text{ (i.e. m = ½)}, \tag{II.2}$$

valid if $d(0)/(R_{T0}+R_W) << 1$.

Derivation of the power dependence of $W_C(0)$ is more cumbersome because $W_C(0)$ depends on d(0) through the angle φ(0). From Eqs.(8) and (10) we obtain:

$$d(0) = (R_{T0} + R_W - R_L) - (C - 2W_C)\frac{1+sin\varphi(0)}{\pi+2\varphi(0)-2cos\varphi(0)}, \tag{II.3}$$

and substitution of $R_T(0)$ (Eq.9) into Eq.(7) leads to:

$$W_C(0) = W_L - 2[R_{T0} + R_W - R_L - d(0)]\frac{cos\varphi(0)}{1+sin\varphi(0)}. \tag{II.4}$$

Since for φ(0) equal to its unloaded value $\varphi_0$, d(0) and $W_C(0)$ are zero, we will suppose that, for small variations δφ around $\varphi_0$, d(0) and $W_C(0)$ are proportional to δφ. Thus we write:

$$\begin{aligned} d(0) &= a\delta\varphi \\ W_C(0) &= b\delta\varphi \\ W_C(0) &= \frac{b}{a}d(0) \end{aligned} \tag{II.5}$$

The proportionality factors *a* and *b* are obtained by taking the first derivative of d(0) and $W_C(0)$ at $\varphi = \varphi_0$ and, if we neglect d(0) in front of $R_{T0} + R_W$, we arrive to their dependence on the wheel geometry:

$$a = (2W_L - C)\frac{cos\varphi_0(\pi+2\varphi_0)-4(1+sin\varphi_0)}{(\pi+2\varphi_0-2cos\varphi_0)^2}$$

$$b = \frac{2acos\varphi_0+2(R_{T0}+R_W-R_L)}{1+sin\varphi_0}. \tag{II.6}$$

Finally, introduction of the $L_C$ and $W_C(0)$ values in the limit of small d(0) (Eqs.(II.6) and (II.9), respectively) into Eq.(22), leads us to the power dependence of F and d(0):

$$F = \alpha P\beta d(0)^{3/2}, \tag{II.7}$$

where $$\beta \equiv \sqrt{2(R_{T0} + R_W)}\left(\frac{b}{a}\right). \tag{II.8}$$

The n = 3/2 exponent is exact if α remains constant during small deformations, and this is true if the shape of the footprint can be parameterized with $W_C(0)$ and $L_C$.

## References

[1] L.D.Landau and E.M.Lifshitz, Course of Theoretical Physics (Pergamon, New York, 1959), vol. 7, sec.9.

[2] A.E.H.Love, *A Treatise of the Mathematical Theory of Elasticity* (Dover, New York, 1927), pp.193-200.

[3] B.Leroy, “Collision between 2 balls accompanied by deformation - a qualitative approach to Hertz theory”. Am.J.Phys. **53**, 346-349 (1985).

[4] Reza H.Jazar, *Vehicle dynamics. Theory and application*. Springer, New York (2008). Chapters 1 and 3. (texkbook for undergraduate students)

[5] Samuel K. Clark, Ed., *Mechanics of Pneumatic Tires* (2nd edition of the original book published in 1971), Department of Transport, Washington (Ref. HS 805 952) (1981).

[6] J.DeEskinazi, W.Soedel and T.Y.Yang,” Contact of an inflated toroidal membrane with a flat surface as an approach to the tire deflection problem”. Tire Sci.Technol. **3**, 43-61 (1975).

[7] A.M.Gent and J.D.Walter, Eds., *The Pneumatic Tire*, Department of Transport, Washington (Ref. HS 810 561) (2006).

[8] J.Sperry and E.Jones, “How do tubeless tires support an auto?”, The Physics Teacher **32**, 174 (1994).

[9] William A.Shurcliff, “What holds up the steel wheels of an automobile?”, The Physics Teacher **32**, 296-297 (1994). (For us, it is not clear that this author give the correct answer)

[10] P.Roura, “Thermodynamic derivations of the mechanical equilibrium conditions for fluid surfaces: Young's and Laplace's equations”. Am.J.Phys. **73**, 1139-1147 (2005).

[11] https://www.exploratorium.edu/snacks/tiredweight

[12] J.T.Tielking, I.K.McIvor and S.K.Clark, “A modified linear membrane theory for the presurized toroid”, J.Appl.Mech. **38** (2), 418-422 (1971).

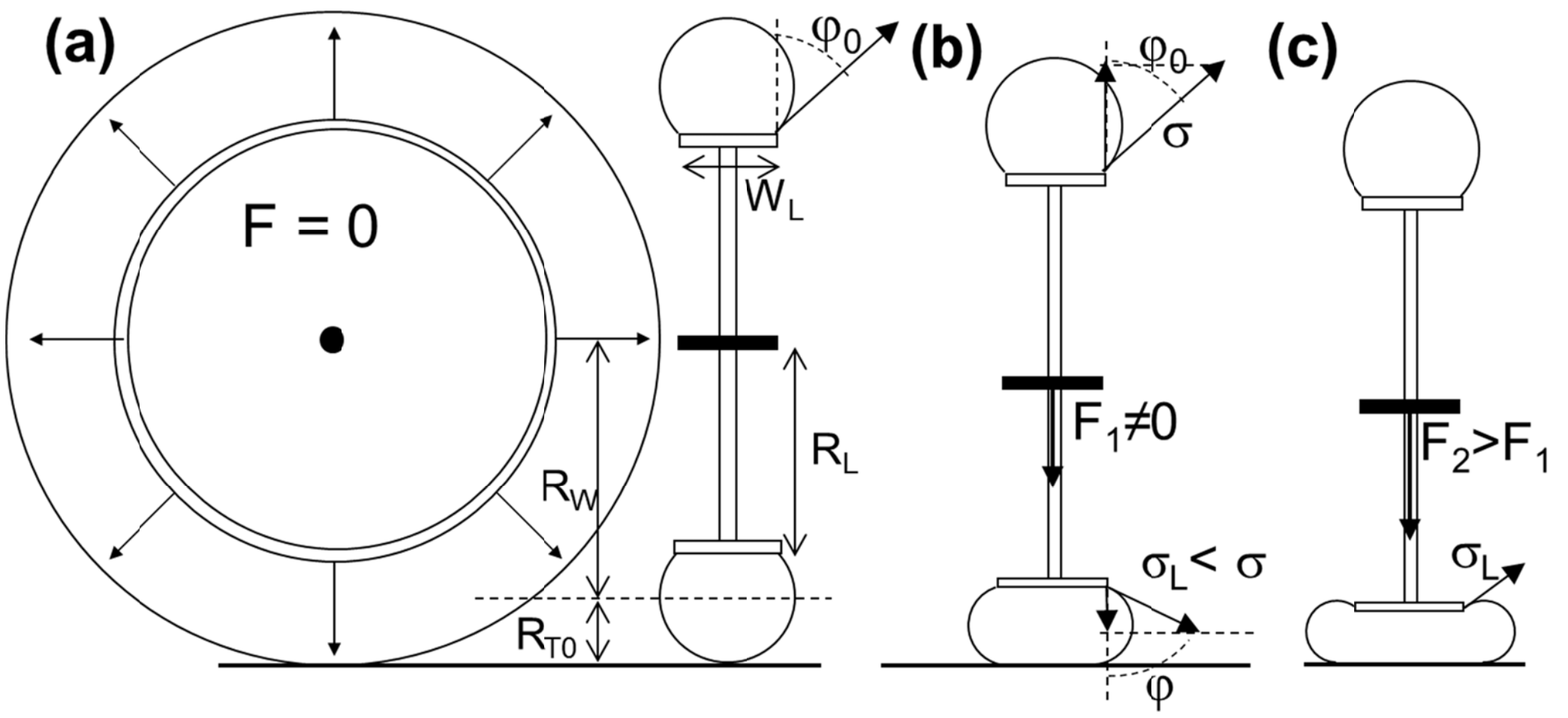

**Figure 1.-** Geometry of a wheel when unloaded (a), loaded and correctly inflated (b) and badly inflated (c) .

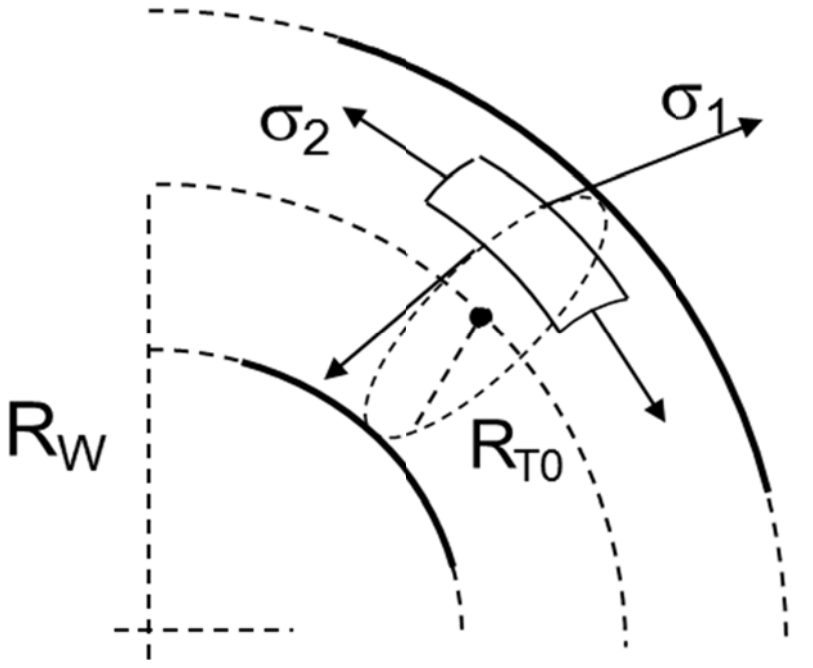

**Figure 2.-** Geometry of a toroid and its principal stresses.

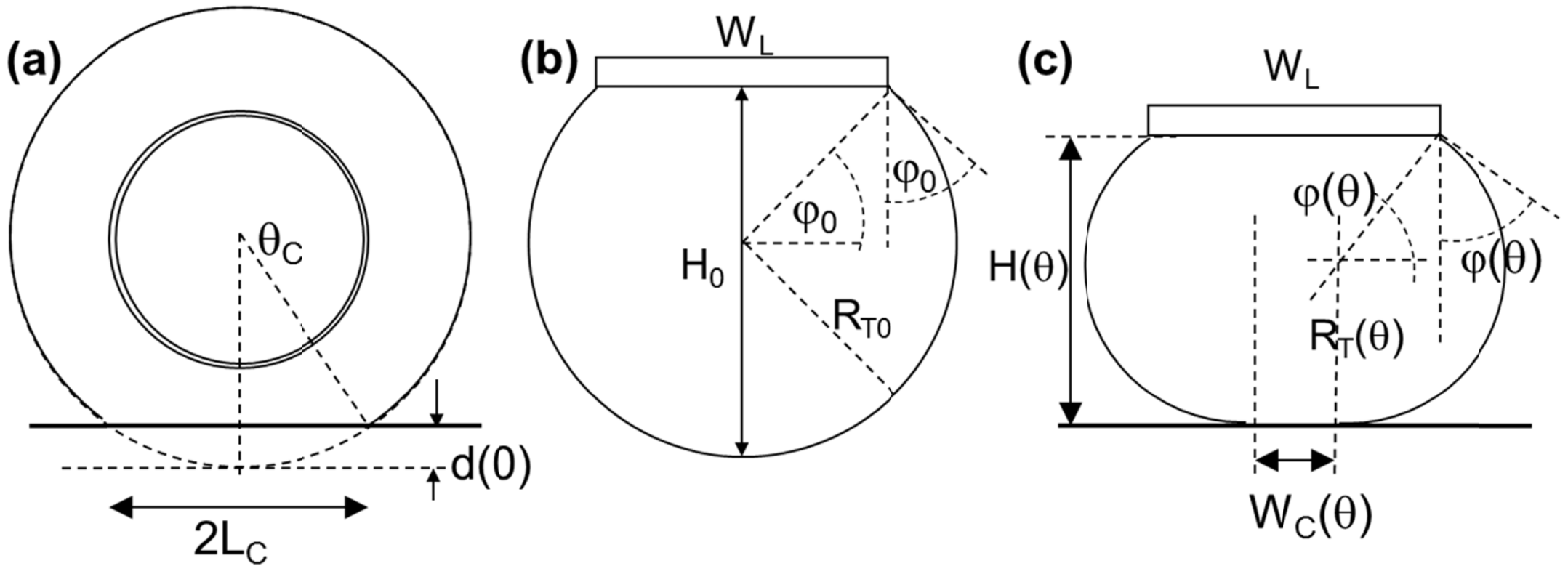

**Figure 3.-** a) Assumption concerning the contact length with the ground of the loaded tire. b) Cross section of the tire when unloaded and for $|\theta| > |\theta_C|$. c) Cross section of the tire for $|\theta| < |\theta_C|$.

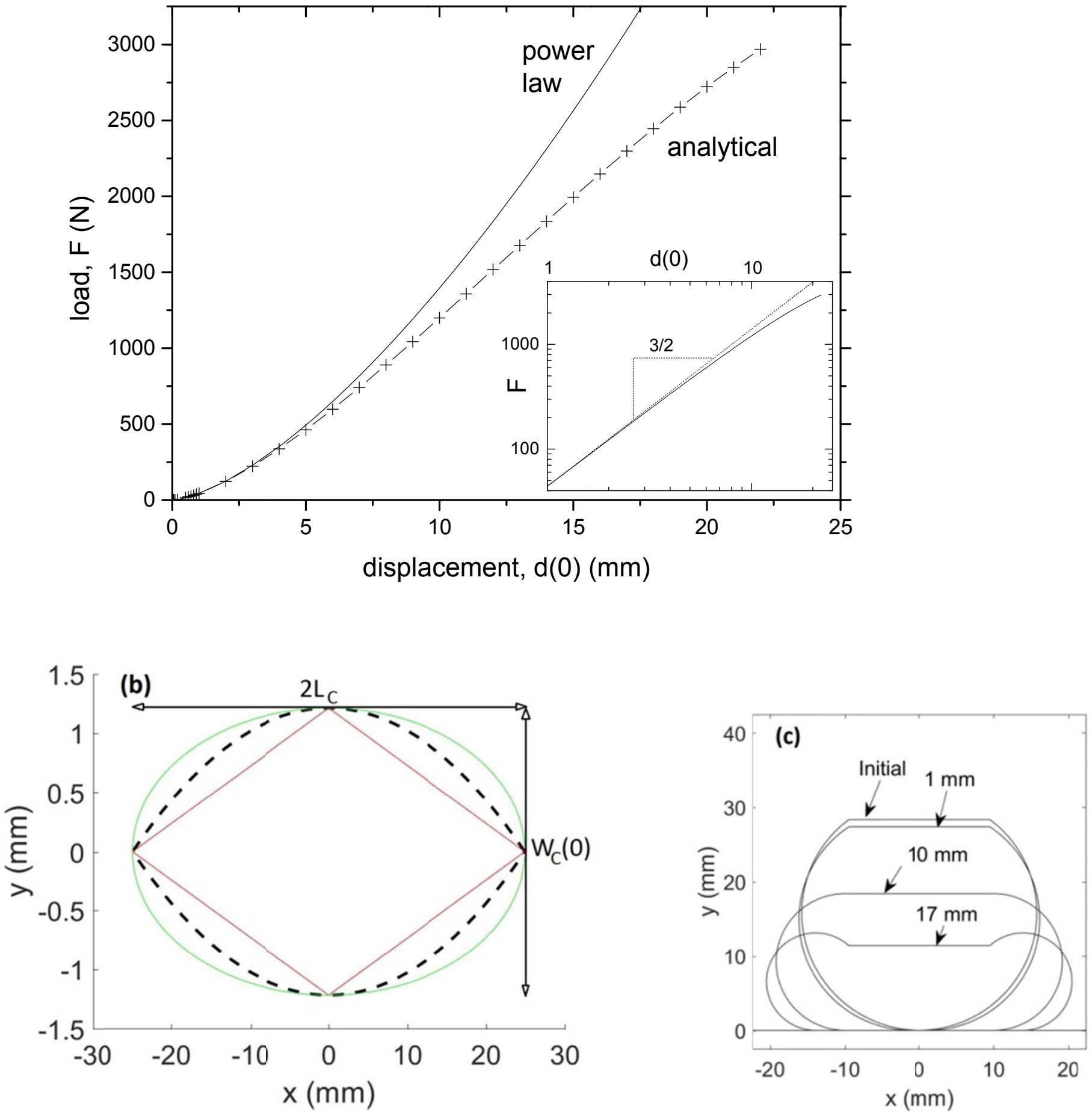

**Figure 4.-** a) Load-deflection curves obtained from the analytical model, and from the power law relationship [Eq.(23)]. Data are from a bike tire inflated at 0.6 MPa. The geometrical parameters are $W_L$= 19.04, $R_{T0}$ = 15.8, $R_W$ = 296.8 and $R_L$ = 282.5mm . The log-log plot of the inset reveals the power dependence with exponent 3/2. b) Footprint (dashed) for a deflection of 1 mm compared with a rhomb and an ellipse. c) Tire cross section in contact with the ground for several loads.

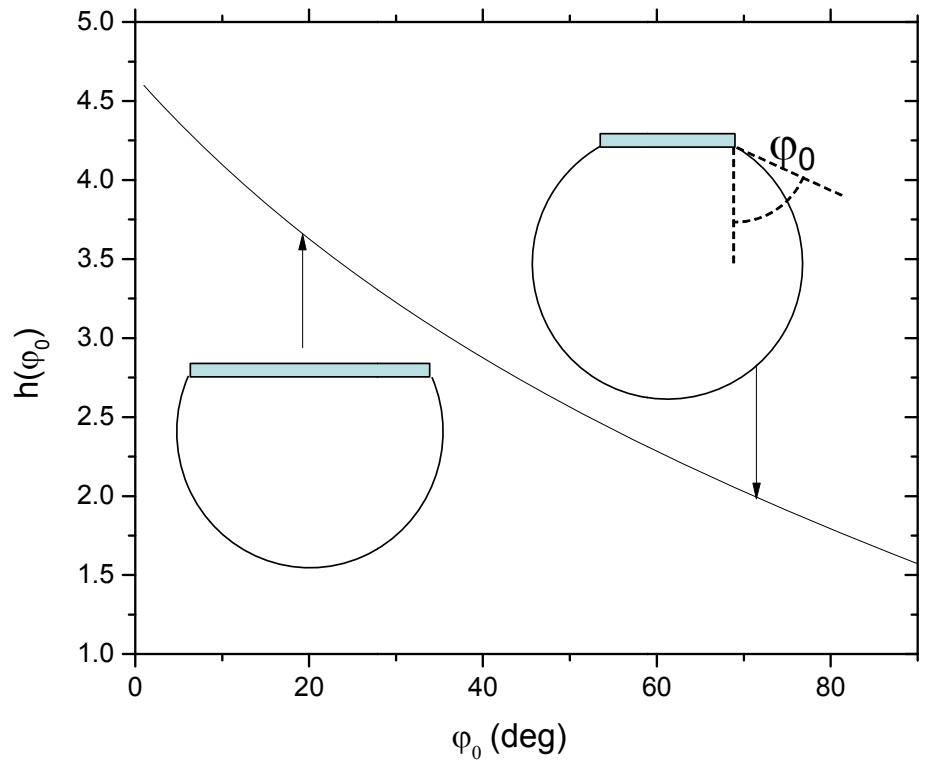

**Figure 5.-** Geometrical factor accounting for the dependence of the force on the tire cross section [Eq.(25)].

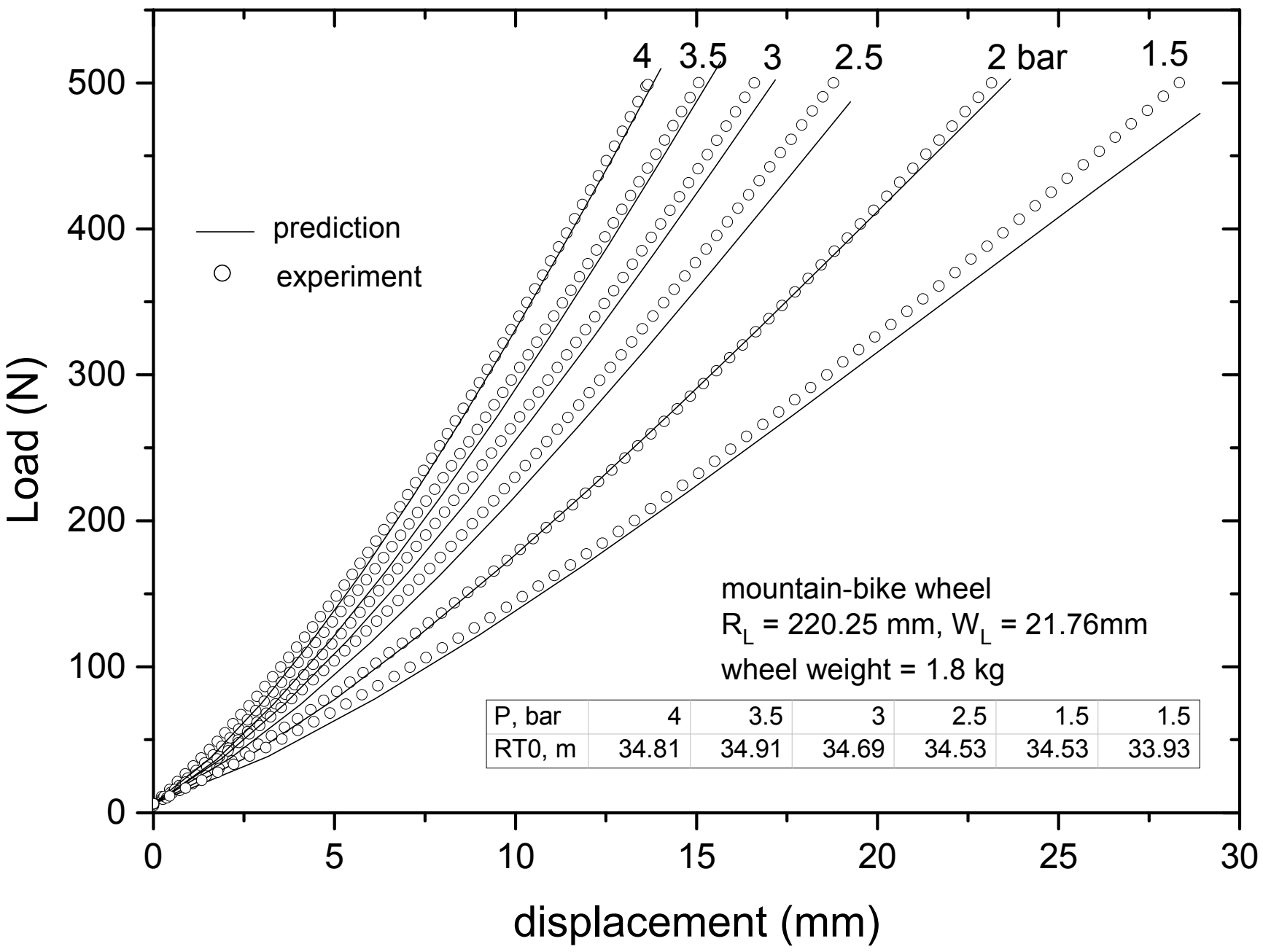

**Figure 6.-** Experimental and predicted load-displacement curves for the mountain-bike wheel measured at several tire pressures. The undeformed cross section and that deformed at 500 N for the lowest pressure are also drawn.

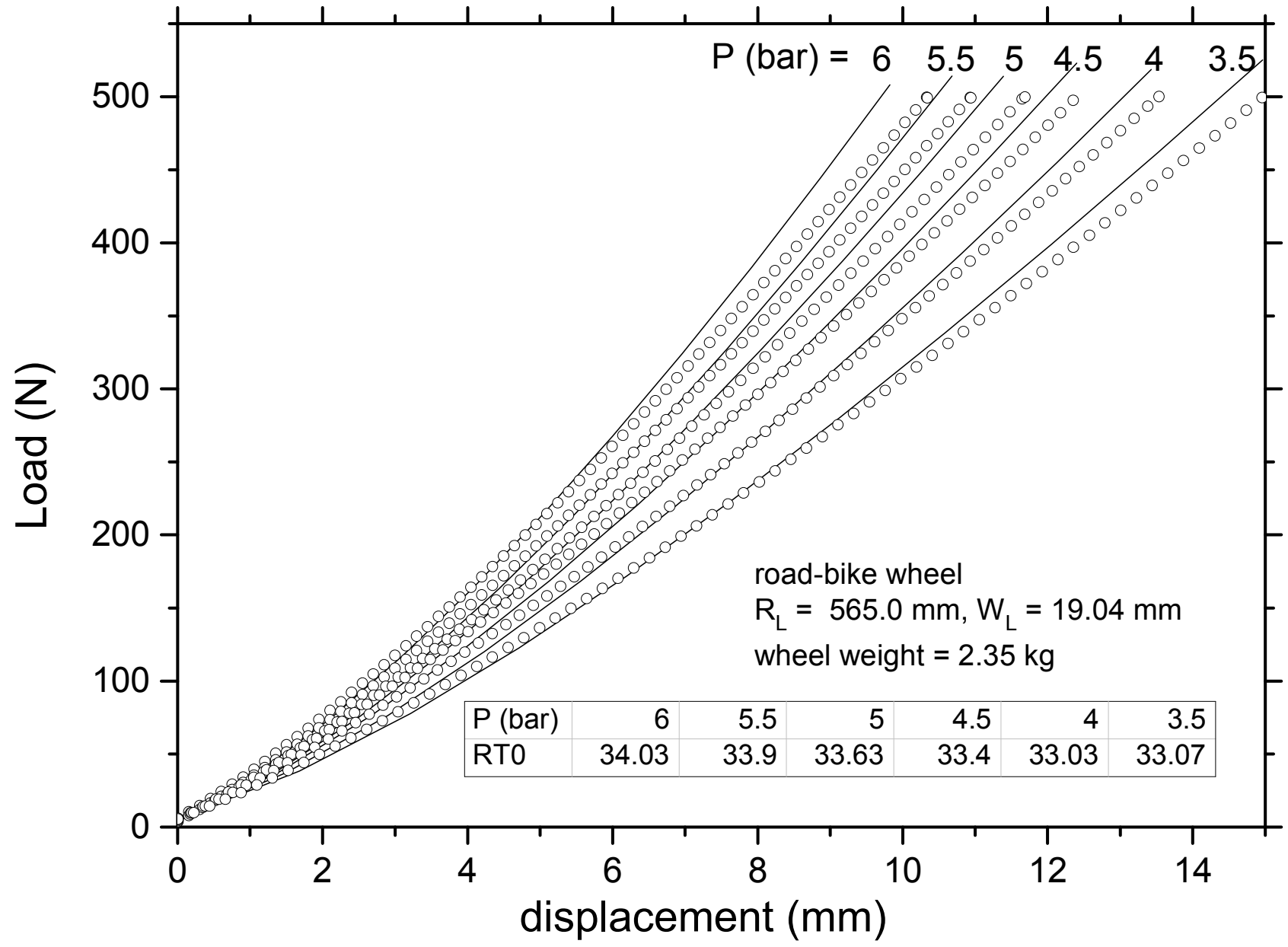

**Figure 7.-** Idem than Fig.6 but for the road-bike.

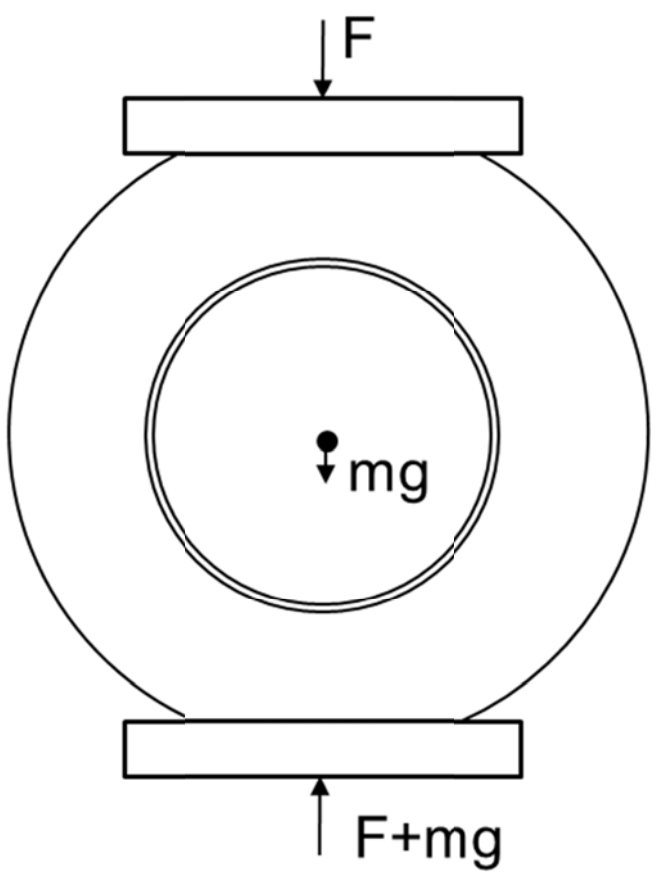

**Figure 8.-** Load, F, applied to the wheels during the experiments (mg is the wheel weight).

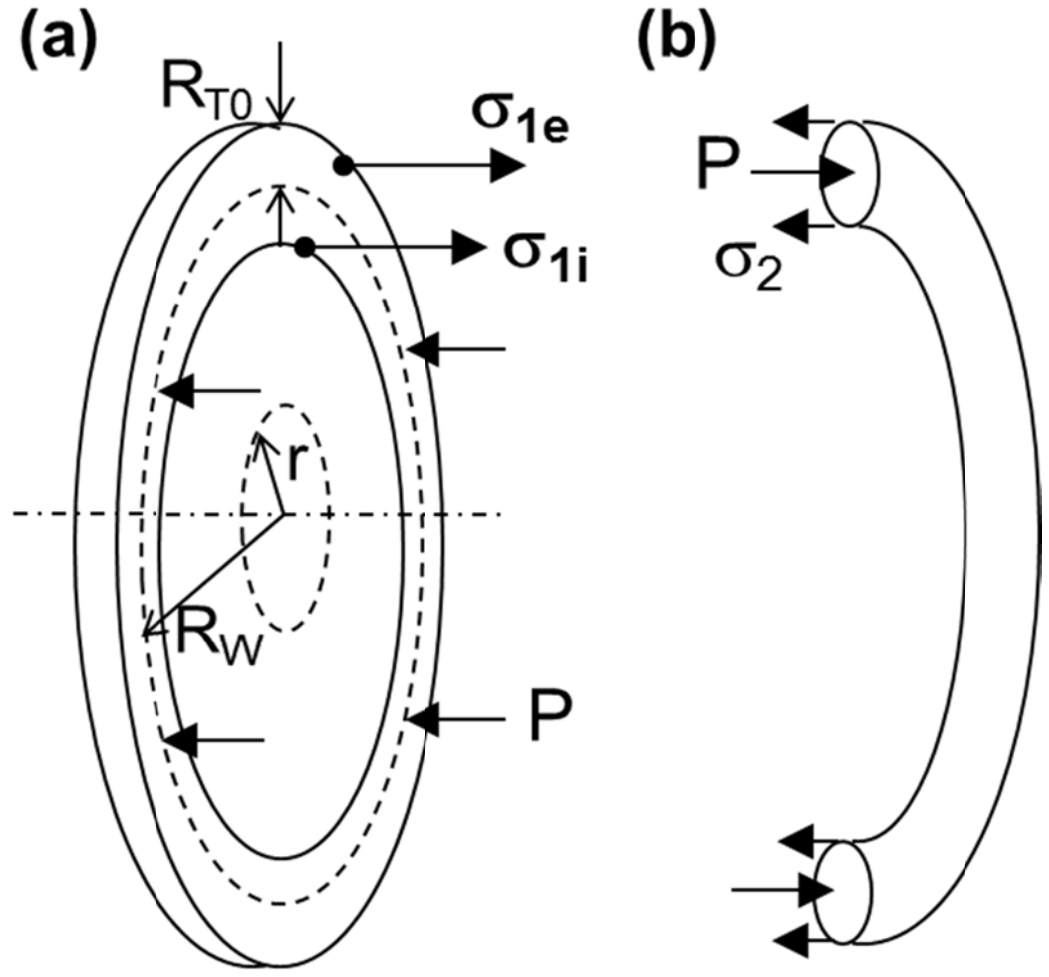

**Figure 9.-** Sections of the toroid used to calculate the tension of the membrane along the principal radii of curvature.